\documentclass{article}
\usepackage{iftex}
\ifXeTeX\usepackage[T1]{fontenc}\fi
\usepackage{microtype}
\usepackage{graphicx}
\usepackage{booktabs}
\usepackage{amsmath,amssymb}
\usepackage{xspace}
\usepackage[hypertexnames=false]{hyperref}
\usepackage{xurl}
\usepackage{pifont}
\newcommand{\stepref}[1]{\ding{\numexpr181+#1\relax}}

\usepackage[accepted]{mlsys2025}
\newcommand{\sys}{\textsc{Weights That Stay}\xspace}
\newcommand{\us}{\,\textmu s\xspace}

\newcommand{\DraftWalkMaxSpeedup}{11.6}

\mlsystitlerunning{Weights That Stay: On-Chip Megakernels for Serial Draft Heads}
\makeatletter
\renewcommand{\Notice@String}{}
\makeatother
\hypersetup{pdftitle={Weights That Stay: On-Chip Megakernels for Serial Heads in Speculative Decoding},pdfauthor={Xinwei Qiang, Ziyi Xu, Zaifeng Pan, Chang Chen, Yufei Ding},pdfsubject={}}

\begin{document}
\twocolumn[
\mlsystitle{Weights That Stay: On-Chip Megakernels\\for Serial Heads in Speculative Decoding}
\begin{mlsysauthorlist}
\mlsysauthor{Xinwei Qiang}{ucsd}
\mlsysauthor{Ziyi Xu}{radix}
\mlsysauthor{Zaifeng Pan}{ucsd}
\mlsysauthor{Chang Chen}{ucsd}
\mlsysauthor{Yufei Ding}{ucsd}
\end{mlsysauthorlist}
\mlsysaffiliation{ucsd}{University of California, San Diego}
\mlsysaffiliation{radix}{RadixArk}
\mlsyskeywords{Speculative decoding, persistent kernels, LLM serving}
\vskip 0.3in
\begin{abstract}
Parallel speculative drafters use a lightweight serial head to recover token dependencies within a block generated
in parallel. Each head step updates only a small state, yet projects that state through the same vocabulary-sized
matrix. Step-wise execution repeatedly transfers this large, immutable projection, making the head a meaningful
source of latency despite its low arithmetic cost.
We present \sys, an execution design that keeps projection shards on chip for the entire serial draft walk.
A persistent megakernel couples weight residency with GPU-side token selection and state advancement, reusing
resident weights while exchanging only compact state between steps. The design supports full or partial residency
and accommodates both Markov and recurrent heads on H100 and B200.
Integrated into SGLang, \sys improves throughput for Markov drafting on Qwen3-4B and Qwen3-8B,
with gains of up to 12.3\%. At the head level, the DSpark walk is up to \DraftWalkMaxSpeedup$\times$ faster than
the fastest measured SGLang reference path. These results show that aligning weight residency with the full draft
walk translates cross-step reuse into end-to-end serving gains.
\end{abstract}
]
\hypersetup{pdftitle={Weights That Stay: On-Chip Megakernels for Serial Heads in Speculative Decoding}}
\printAffiliationsAndNotice{}
\section{Introduction}\label{sec:intro}

Speculative decoding accelerates language-model generation by drafting multiple tokens and verifying them in
parallel with the target model~\citep{leviathan2023fast,chen2023accelerating}.
Parallel drafters produce a block in one backbone pass, but predictions made without their realized predecessors
can lose important intra-block dependencies. DSpark~\citep{cheng2026dspark} and Domino~\citep{huang2026domino}
recover these dependencies with a lightweight serial head after the parallel backbone. Each head step uses tokens selected
earlier in the block to refine the next proposal.

The serial head performs little arithmetic, but can still add substantial latency.
Although its projection is low-rank, it spans more than 150,000 vocabulary rows and is larger than the H100 L2
cache. Reapplying this projection at every dependent step makes the head bandwidth-oriented at small batch. In
SGLang, the head accounts for up to 12\% of a Qwen3-4B speculative round. Low rank reduces the projection's
footprint, but does not eliminate its repeated movement along the serial path.

\begin{figure}[t]
\centering
\includegraphics[width=\linewidth]{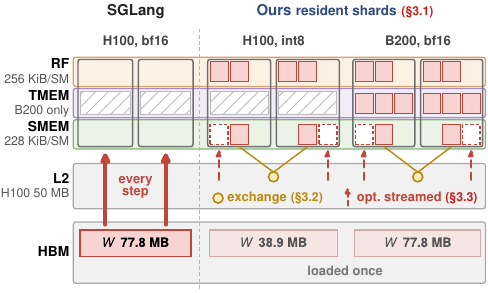}
\caption{Walk-resident execution across GPU memory tiers. Each SM retains a shard of $W$ throughout the walk,
exchanging only tokens or recurrent state. Weights are int8 on H100 and bf16 on B200; optional loads stream
nonresident tiles.}
\label{fig:overview}
\end{figure}

The root cause is a mismatch between the execution boundary and the lifetime of reusable data.
The vocabulary projection is unchanged across the entire walk, while the cross-step dependency is carried by a
much smaller token or recurrent state. Conventional operator boundaries confine the projection's
software-managed lifetime to one operation or step: registers and shared memory do not persist across kernel
launches.

The same projection is used at every step, so its shards can remain resident throughout the walk. A whole-walk
kernel exposes this opportunity but does not realize it automatically: it can still reload the same weights at
every iteration. We therefore assign fixed row shards to a co-resident grid for the complete walk. Each CTA loads
retained weight tiles once, scores the same vocabulary rows at every step, and exchanges only the compact
information needed to advance the recurrence. This decouples traffic for the resident weights from the number of
steps in the walk.

This organization creates three coupled design requirements. First, the projection needs a stable mapping at two
levels: vocabulary rows to SMs, and each SM's shard to the storage locations consumed by its matrix instructions.
Second, global token selection and state advancement must complete without relinquishing that mapping. Third,
the design must exploit partial fit when a shard modestly exceeds the operand storage left after reserving the
execution workspace.

We realize this design in \sys, a family of persistent megakernels for serial draft heads
(Figure~\ref{fig:overview}). A GPU-side collective advances the recurrence while preserving fixed shard ownership.
Resource-specific layouts provide full or partial residency on H100 and B200. The same execution structure
supports heads with Markov, RNN, and GRU-based state updates.

On H100, the DSpark walk is up to \DraftWalkMaxSpeedup$\times$ faster than the fastest measured SGLang reference
path. With block-16 Markov drafting, SGLang client throughput improves by up to 12.3\% for Qwen3-4B and 8.6\%
for Qwen3-8B. Head-level results extend across Markov and recurrent heads and from H100 to B200.

This paper makes three contributions:
\begin{itemize}
\item We characterize serial draft heads by a large immutable projection shared across steps and a small evolving
state, and show why step-wise execution fails to exploit this temporal reuse (Section~\ref{sec:workload}).
\item We develop walk-resident execution as a hierarchical tile-mapping design space, together with a GPU-side
protocol that advances dependent steps without breaking the mapping (Section~\ref{sec:design}).
\item We implement the design on H100 and B200, achieving up to \DraftWalkMaxSpeedup$\times$ head-level speedup
across Markov and recurrent heads and 12.3\% higher SGLang client throughput for Markov drafting
(Sections~\ref{sec:implementation}--\ref{sec:eval}).
\end{itemize}

\section{Serial Draft Heads and Their Execution Mismatch}\label{sec:workload}

\subsection{Speculative Decoding with Serial Draft Heads}\label{sec:spec-decoding}
Speculative decoding accelerates autoregressive generation by using a lightweight drafter to propose multiple
future tokens and the target model to verify them in parallel~\citep{leviathan2023fast,chen2023accelerating}.
In each round, the drafter proposes a token block and the target scores its positions in one pass.
Greedy verification
accepts the longest prefix that agrees with the target's choices. For sampling, a proposed token $y$ drawn from
the draft distribution $q$ is accepted with probability $\min(1,p(y)/q(y))$, where $p$ is the target distribution.
On rejection, the replacement is sampled from a distribution proportional to $\max(p-q,0)$. This correction
preserves the target distribution as long as drafting and verification use the same $q$.

Drafter quality therefore affects efficiency rather than target correctness. Better proposals produce longer
accepted prefixes, allowing each target pass to advance generation farther. Drafter latency and accepted length
together determine the end-to-end benefit.

Parallel drafters reduce drafting latency by computing the base logits and slot features for all draft positions
in one backbone forward pass. Because the positions are computed together, their base predictions cannot depend
on the tokens actually selected at earlier positions in the block. The systems we study add a serial correction
head. After selecting token $y_k$, the head carries that token or a recurrent state
into the next step to refine the next proposal. This restores intra-block dependence, but places a dependent
head walk after the parallel backbone.

\begin{figure}[t]
\centering
\includegraphics[width=\linewidth]{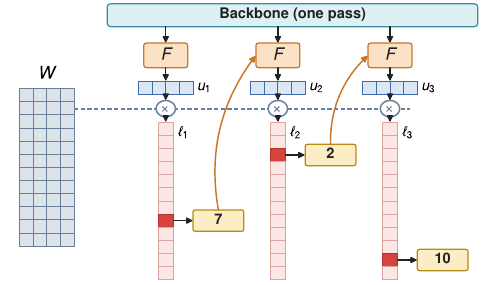}
\caption{Serial refinement with a shared vocabulary projection ($V=12$, $r=4$, three steps). Each selected
token advances the next step. Base-logit addition and explicit state-update nodes are omitted; the full
recurrence is given in Equations~\ref{eq:cell}--\ref{eq:advance}.}
\label{fig:background}
\end{figure}

We use $\gamma$ for the number of serial head steps executed in a walk, which can differ from the checkpoint's
nominal block size. Denote the backbone outputs for step $k$ by base logits
$b_k\in\mathbb{R}^{V}$ and slot features $x_k$, for $k=1,\ldots,\gamma$. The relevant heads can be expressed as
\begin{align}
u_k &= F(s_{k-1},x_k), \label{eq:cell}\\
\ell_k &= b_k + W u_k, \label{eq:projection}\\
y_k &= \operatorname{Select}(\ell_k), \label{eq:select}\\
s_k &= G(s_{k-1},x_k,y_k). \label{eq:advance}
\end{align}
The vocabulary projection $W\in\mathbb{R}^{V\times r}$ is fixed across steps (Figure~\ref{fig:background}).
The latent $u_k\in\mathbb{R}^{r}$ and state $s_k$ depend on the generated prefix.
Selection is an argmax for greedy decoding or a sample from the temperature-scaled proposal distribution.

For DSpark's Markov head, $s_{k-1}$ is the previous token and $F$ gathers its row from an embedding table.
Domino~\citep{huang2026domino} and DeLS-Spec~\citep{zheng2026dels} maintain recurrent state to summarize more of
the prefix before the low-rank projection.
Their state updates differ in how they compute $u_k$, but retain the repeated vocabulary-wide operation in
Equation~\ref{eq:projection}. We target this shared structure. Purely parallel heads and methods whose serial
phase is already confined to a small candidate lattice occupy a different regime.

\subsection{Why a Lightweight Head Is Still Expensive}
These heads are small relative to the backbone, but conventional execution reads their vocabulary projection
once for every dependent step. At batch $B$, the projection performs approximately $2BVr$ floating-point
operations per step. If a weight uses $p$ bytes, reading the table transfers $Vrp$ bytes per step and
$\gamma Vrp$ bytes over the walk.
Ignoring other traffic, the arithmetic intensity is therefore about $2B/p$ operations per byte.
At small batch, this is a bandwidth-oriented operation despite its large output dimension.
The dependency in Equation~\ref{eq:advance} prevents evaluating the realized sequence of latents independently.

\begin{figure*}[t]
\centering
\includegraphics[width=\textwidth]{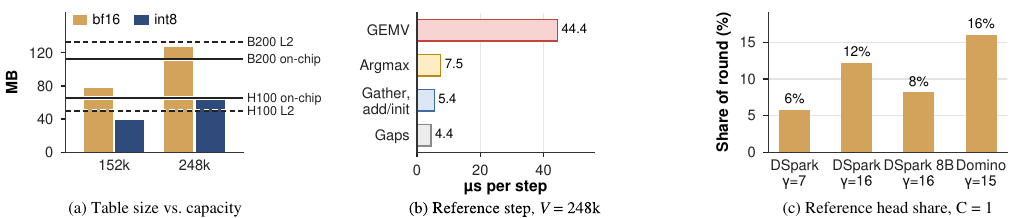}
\caption{Memory footprint and latency cost of serial draft heads. (a) Rank-256 projection size versus nominal
GPU capacities (MB). (b) One SGLang step at $B=1$, $V=248$k under CUDA-graph replay.
(c) Head microbenchmark latency divided by measured serving-round latency at $C=1$, $T=0$;
DSpark uses the default folded path, and Domino uses the full-vocabulary head.}
\label{fig:motivation}
\end{figure*}

With $r=256$, the bf16 projection occupies 77.8\,MB at $V=151{,}936$ and 127.1\,MB at $V=248{,}320$.
Both exceed H100 L2 capacity (Figure~\ref{fig:motivation}a). A conventional walk repeatedly streams the projection through the memory hierarchy,
with additional gather, epilogue, and selection operations at every step.
A profile of the larger-vocabulary SGLang head shows that the vocabulary projection takes 44.4\us per step and
dominates its latency (Figure~\ref{fig:motivation}b).

The round-level impact depends on the workload (Figure~\ref{fig:motivation}c). Longer walks repeat the head cost more often; larger targets
increase verification time and dilute the same saving. Larger batches also increase target compute and change
reuse across requests. Our evaluation therefore varies walk length, target size, and batch size, and reports
both head-level and serving-level effects.

\subsection{Aligning Execution with Reuse}
The analysis above exposes a mismatch between algorithmic reuse and execution lifetime: the projection weights
remain unchanged for the entire walk, but conventional operator execution retains each tile in registers or
shared memory for only one projection. The next dependent step must stage the same tile again from L2 or HBM
for a new latent.

Walk-resident execution reverses this dataflow. A persistent CTA retains a weight shard and brings each successive
latent to it. The recurrence remains serial, but the dominant weights now outlive the step boundary.
Section~\ref{sec:design} develops the required shard mapping, state-advance protocol, and partial-residency path.

\section{Walk-Resident Execution}\label{sec:design}
Extending weight lifetime across a walk creates three problems: mapping the projection across SMs and their
storage tiers, advancing a distributed recurrence without losing that mapping, and handling shards that do not
fit alongside the execution workspace.

\subsection{A Hierarchical Tile-Mapping Space}\label{sec:organization}
\begin{figure}[t]
\centering
\includegraphics[width=\linewidth]{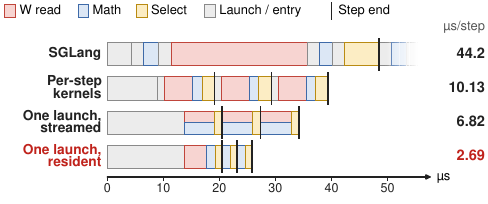}
\caption{Whole-walk fusion removes launch boundaries, while residency removes repeated fetches of $W$ from L2 or HBM
($V$=151,936, $B$=1, greedy).
Timings are fitted from the complete-call measurements in Section~\ref{sec:ablation}.
The SGLang row is truncated after one step, and the within-step segments are illustrative.
The resident kernel's entry-cost excess over streaming is drawn as a one-time shard load.}
\label{fig:strategies}
\end{figure}

Per-step fusion reduces intermediate traffic and operator launches within each step.
Fusing the entire walk further removes kernel boundaries between steps.
The kernel can still repeat the same weight transfers at every iteration.
We retain tiles across steps to reduce this recurring copy work.
Figure~\ref{fig:strategies} shows the weight reads and execution boundaries in these strategies.

Residency requires a stable mapping at two levels. Across SMs, we partition $W$ by vocabulary rows and assign one
shard to a persistent CTA on each participating SM. A CTA owns the same token rows, spanning the full latent
dimension, for the complete walk. Within an SM, the shard is divided among storage tiers, with instruction-specific
paths to supply operands. Both mappings remain fixed while the latent and selected token change
between steps.

\begin{figure}[t]
\centering
\includegraphics[width=\linewidth]{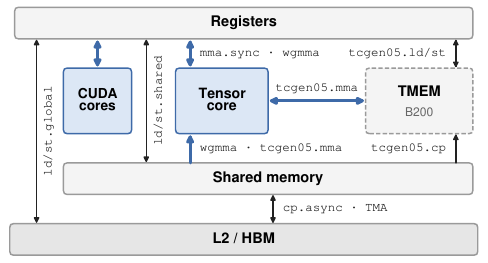}
\caption{Operand storage and data paths for GPU matrix instructions~\citep{nvidia_ptx}. Thin arrows denote
copies; thick arrows denote operand and result access. \texttt{mma.sync} uses registers; \texttt{wgmma} reads
$A$ from RF/SMEM and $B$ from SMEM, accumulating in RF; \texttt{tcgen05.mma} reads $A$ from TMEM/SMEM and
$B$ from SMEM, accumulating in TMEM.}
\label{fig:datapath}
\end{figure}

Together, these choices define a hierarchical tile-mapping space. For each weight tile, a layout chooses its
persistent SM owner, its storage tier within that SM, and the matrix instruction that consumes it.
Figure~\ref{fig:datapath} distinguishes copies between storage tiers from operand accesses by computation
instructions. Copy instructions move tiles between tiers, whereas matrix instructions consume operands from
prescribed locations~\citep{nvidia_ptx}. A tile is resident when it remains in software-managed on-chip storage
across dependent steps, avoiding repeated fetches from L2 or HBM. Residency does not eliminate all on-chip
movement: shared-memory tiles may still need to be loaded into registers before computation.

The mapping therefore determines both the resident capacity and the recurring operand-staging cost. Making
another tile resident removes its repeated fetch from L2 or HBM but consumes on-chip storage. Changing the matrix
instruction changes the operand locations and staging paths, affecting both usable capacity and per-step cost.

\subsection{A Distributed State-Advance Protocol}\label{sec:advancement}
Fixed ownership distributes the projection: CTA $j$ produces scores only for the rows in its shard $W_j$. The
recurrence, however, has one logical state, and no CTA can begin the next step until that state reflects a token
selected from the full vocabulary. We therefore execute each step as a distributed state transition with four
ordered phases: prepare, score, select, and commit.

In the prepare phase, the head-specific update exposes the latent components needed by every weight owner. Each
CTA then scores its local rows without communication. The select phase computes one token from the distributed
scores and publishes it to every CTA, using an argmax for greedy decoding or categorical selection for sampling.
Finally, the commit phase applies the head-specific state update and establishes the state for the next step. No
CTA enters the next prepare phase before commit completes. Weight ownership remains unchanged across all four
phases. Only the selected token and required state fragments cross CTA boundaries.

This protocol permits different state layouts. A Markov head needs no state exchange beyond selection: every CTA
uses the common token to gather the next latent locally. A replicated recurrent update is likewise recomputed by
every CTA, while a sharded update adds collectives for the required state fragments during prepare or commit.
Section~\ref{sec:realization-cells} gives the concrete mappings, and Section~\ref{sec:head-generality} measures
their critical paths.

The collectives impose a progress condition. If waiting CTAs occupy resources needed by CTAs that have not been
scheduled, the grid can deadlock. We therefore use a cooperative launch with one CTA on each participating SM so
that every weight owner remains live for the complete walk. Section~\ref{sec:realization-exchange} describes the
implemented exchange.

\subsection{When the Shard Does Not Fit}\label{sec:resources}\label{sec:placement}\label{sec:configuration}
The mapping in Section~\ref{sec:organization} must share each SM with the execution workspace: accumulators, the
current latent and state, selection scratch space, and buffers for data movement. After reserving this
workspace, the remaining operand storage may not hold the complete weight shard. Larger vocabularies increase the
shard, while larger batches and more complex state updates increase the workspace.

We use two complementary responses to this capacity gap. A compact weight representation reduces the bytes in
every row and can make the complete shard resident. If the shard still does not fit, the layout retains a fixed
subset of tiles and streams the rest from L2 or HBM at every step. These choices can be combined: representation
determines the shard size, while the resident fraction determines how much recurring traffic remains.

Resident tiles, staging buffers, and execution state compete for on-chip storage. A larger resident subset removes
recurring reads but leaves less buffering to overlap streamed transfers or less workspace for batched execution
and state.
Section~\ref{sec:implementation} instantiates these choices on H100 and B200.

\section{GPU Kernel Realization}\label{sec:implementation}
We instantiate walk-resident execution by mapping each weight shard to GPU operand storage and advancing the
recurrence without relinquishing that mapping.

\subsection{Weight Placement and Execution}\label{sec:layout-realization}
The mapping in Section~\ref{sec:organization} must satisfy the operand locations of its matrix instructions after
reserving the workspace identified in Section~\ref{sec:resources}. Each implementation therefore chooses a weight
representation, an on-chip placement, and a schedule for any tiles that remain outside that placement.
The resulting paths are summarized below.

\begin{center}
\centering
\small
\begin{tabular}{@{}lrrrr@{}}
\toprule
 & \multicolumn{3}{c}{Resident rows per CTA} & Streamed \\
\cmidrule(lr){2-4}
Case & RF & SMEM & TMEM & per step \\
\midrule
H100 int8, small $B$ & 768 & 384 & -- & -- \\
H100 int8, large $B$ & -- & 704 & -- & 448 \\
H100 int8, large $V$ & -- & 704 & -- & 1,216 \\
B200 bf16, standard $V$ & 336 & 224 & 480 & -- \\
B200 bf16, large $V$ & 336 & 264 & 440 & 640 \\
\bottomrule
\end{tabular}
\end{center}

\begin{figure}[t]
\centering
\includegraphics[width=\linewidth]{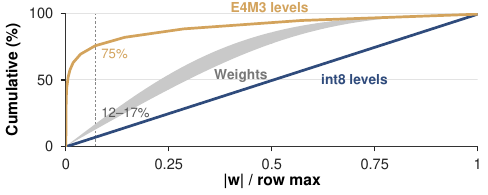}
\caption{Cumulative distributions of normalized weight magnitudes (band: two released 4B heads) and
representable positive values (quantization levels). Left of the dashed line, E4M3 has finer spacing than int8
but places 75\% of its levels in a region containing only 12--17\% of the weights.}
\label{fig:quant}
\end{figure}

\paragraph{H100 weight representation and execution.}\label{sec:h100-single}
The standard bf16 projection occupies 77.8\,MB, already exceeding H100's 65.4\,MB aggregate register and
shared-memory capacity before reserving execution workspace. Full residency therefore requires a one-byte weight
format. Both int8 and FP8 provide this footprint, but allocate their bits differently. After per-row scaling, the
inspected weight rows are dense and approximately bell-shaped, without the heavy tails that would require a wide
exponent range (Figure~\ref{fig:quant}). Int8 distributes its levels uniformly across the bounded row range, whereas FP8 E4M3 spends bits
on exponent range and has coarser spacing at the larger magnitudes that contribute most to projection error. We
therefore store $W$ in row-scaled int8, reducing the table to approximately 38.9\,MB before scale metadata and
making full residency feasible.

\begin{figure}[t]
\centering
\includegraphics[width=\linewidth]{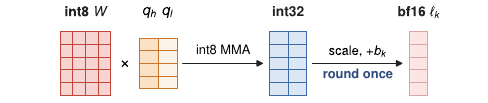}
\caption{Vocabulary projection using int8 tensor cores on H100. Two int8 vectors represent the latent $u$
and are multiplied by $W$ together. Their scaled results are summed, added to the base logits, and rounded
once to bf16.}
\label{fig:numerics}
\end{figure}

To avoid dequantizing the resident weights at every step, we execute the projection with native int8 tensor-core
instructions. These instructions require the latent operand to be int8 as well. Each request moves only one rank-256
latent per step, so carrying a second int8 vector adds little traffic. We spend this small extra traffic to retain
more latent precision, encoding $u$ in two int8 byte planes. The high plane $q_h$ encodes the coarse value, while
the low plane $q_l$ encodes its residual at one 256th of the high-plane scale,
\begin{equation}
u\simeq s_h q_h+s_l q_l,\qquad s_l=s_h/256.
\end{equation}
For a weight row $W_i\simeq a_i q_i$, the projection becomes
\begin{equation}
W_i u\simeq a_i\bigl(s_h\langle q_i,q_h\rangle+s_l\langle q_i,q_l\rangle\bigr).
\end{equation}
Both dot products now use int8 tensor cores without expanding the weight tiles to floating point at each step.
We place $q_h$ and $q_l$ in adjacent columns of the MMA operand, allowing the kernel to compute their projections
together. Each request contributes one such column pair. The int8 dot products accumulate in int32, after which
the epilogue applies the two scales, sums the results, and adds the base logits (Figure~\ref{fig:numerics}).

The resulting int8 operand tiles are split between registers and shared memory within each CTA's fixed shard.
Register-resident tiles remain in \texttt{mma.sync} operand fragments; shared-memory tiles are loaded into
operand registers when used. Both remain on chip across steps. The table above gives the row allocation
after reserving execution workspace.

\paragraph{Partial residency on H100.}\label{sec:h100-batched}
When the H100 shard does not fit for a given batch size and vocabulary size, the implementation selects a
partial-residency kernel. Both its resident weights and the tiles delivered by TMA reside in shared memory, so
the kernel uses \texttt{wgmma} to consume them there directly while keeping the latent in registers. Shared memory is divided
between a fixed set of resident tiles and staging space that TMA refills at each step; both sets of tiles then
follow the same \texttt{wgmma} path.

\paragraph{B200 layouts.}\label{sec:b200-realization}
B200 adds tensor memory (TMEM) as an on-chip operand store accessed through the \texttt{tcgen05} instruction
family. Combining TMEM with register and shared-memory placement provides enough operand capacity to keep the
standard-vocabulary projection in bf16. We divide each CTA's row shard among these three locations:
TMEM-resident tiles use \texttt{tcgen05}, while the other tiles use \texttt{mma.sync}. The ownership of vocabulary
rows remains unchanged.

For the larger vocabulary, B200 uses the same partial-residency organization as H100: a fixed subset of bf16
tiles remains on chip, while TMA stages the rest in shared memory at each step.

\subsection{Selection and State Advancement}\label{sec:state-realization}
The projection leaves each CTA with scores only for the vocabulary rows it owns. Advancing the walk requires all
CTAs to agree on one token and then materialize the state or latent consumed by the next projection.

\begin{figure}[t]
\centering
\includegraphics[width=\linewidth]{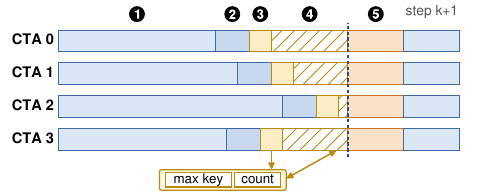}
\caption{Cross-CTA token selection. Each CTA scores its rows \stepref{1}, packs its local
winner \stepref{2}, atomically updates the maximum key and arrival count \stepref{3}, waits for all CTAs
\stepref{4}, and uses the common winner \stepref{5}. The \{key, count\} observation uses the hardware assumption
stated in the text.}
\label{fig:advance}
\end{figure}

\paragraph{Cross-CTA token selection.}\label{sec:realization-exchange}
Each CTA first reduces its rows to a local score and token identifier. It packs them into a 64-bit key whose score
bits preserve numerical order and whose inverted token bits break ties toward the smaller identifier. An integer
maximum over these keys therefore implements the desired global selection. One thread in each CTA atomically
updates the shared maximum and then atomically increments an arrival counter (Figure~\ref{fig:advance}). The
increment uses release ordering so that the key update precedes the CTA's arrival. The cooperative grid keeps
every contributing CTA resident while they wait.

After contributing, each CTA waits until the arrival count reaches the number of participating CTAs and then
reads the maximum key. Our H100 and B200 kernels combine these two operations by polling the aligned 16-byte
\{key, count\} pair with one relaxed vector load. When a load returns the final count, the CTA uses the key
returned by the same load, avoiding a dependent L2 round trip. This relies on behavior observed on both GPUs: the
aligned load sees both fields from one L2 snapshot. PTX~\citep{nvidia_ptx} does not guarantee this property for
vector loads. A portable implementation can instead poll the count with volatile loads until completion,
execute an acquire fence, and then load the key separately. This increases measured walk latency by 6.7\%.

\paragraph{Greedy and stochastic selection.}
The final bf16 logits from the epilogue define the proposal distribution used by both selection and verification.
Greedy requests pack these logits directly into the keys above.

For stochastic selection, we use two standard implementations of categorical sampling. Inverse-CDF sampling
draws one uniform variate and locates it in the cumulative probability mass. Gumbel-max instead adds independent
Gumbel noise to each log mass and selects the maximum, turning categorical sampling into a reduction. Batched
sampling combines the two methods in two levels. Each eight-row group first produces a sample conditioned
on that group, and a second sample selects one group according to its total probability mass. At the first level,
one thread holds all eight logits, so inverse CDF needs only one uniform variate and a register-local prefix sum.
Gumbel-max would generate and transform one variate per row. At the second level, group masses are distributed
across threads and CTAs, so inverse CDF would require a global prefix scan. Gumbel-max over the log masses instead
uses the same maximum reduction as greedy selection. The selected group returns its conditional sample, yielding
a sample from the original categorical distribution. Philox~\citep{salmon2011philox} counters key the noise by
request, vocabulary row, step, and graph replay, so repeated graph executions draw fresh samples.

\paragraph{From the selected token to the next latent.}\label{sec:realization-cells}
A Markov head maps the selected token directly to the next latent, so token selection is its only collective.
DeLS-Spec and Domino instead follow
\begin{equation}
\begin{aligned}
h_k &= \mathrm{GRU}(E[y_{k-1}],h_{k-1}),\\
u_k &= P_k(h_k), \qquad \ell_k=b_k+Wu_k.
\end{aligned}
\end{equation}
Here $h_k$ is the recurrent history, $u_k$ is the low-rank input to the vocabulary projection $W$, and $P_k$
denotes the head-specific projection from one to the other.

We precompute the GRU input contribution for every token and shard the recurrent matrix across CTAs. While
token selection is in flight, each CTA computes the part that depends only on the previous hidden state. Once the
token is known, it looks up the corresponding precomputed input and finishes its slice of the new hidden state.
The following latent projection needs the complete hidden state, so the CTAs write their slices to a
global-memory buffer and wait until all contributors have arrived. Every CTA then reads the complete hidden state.

The new hidden state must then be mapped to the latent consumed by the vocabulary projection. The matrix in $P_k$
is also row-sharded, so each CTA produces only a slice. Every vocabulary-weight owner needs the complete latent to
score its rows, so the CTAs write their latent slices to the buffer and synchronize a second time. Every weight
owner then reads and quantizes the complete latent. The first gather therefore supplies the hidden state to the
latent projection, and the second supplies the latent to the vocabulary projection. When an update matrix is small
enough to replicate in every CTA, each CTA can compute the complete latent locally and needs no gather. The
DSpark RNN instead produces latent slices and gathers them once.
Figure~\ref{fig:heads}b separates this state-update cost from the shared projection.

In SGLang~\citep{zheng2024sglang}, the H100 Markov kernel replaces only the serial draft walk within the existing
draft-and-verify round; backbone execution and target verification remain unchanged.

\section{Evaluation}\label{sec:eval}

\subsection{Experimental Setup}
\paragraph{Platforms.}
We use NVIDIA H100 SXM GPUs (80\,GB, 132 SMs) and B200 GPUs (148 SMs) with CUDA 13. H100 is the primary
platform, and each serving configuration runs on a single H100. The B200 experiments evaluate head execution only,
with reference and optimized paths measured in separate B200 containers.

\paragraph{Workload.}
The main serving experiments use 3,030 prompts from nine math, code, and conversation datasets:
GSM8K~\citep{cobbe2021trainingverifierssolvemath}, MATH500~\citep{lightman2023lets},
AIME25~\citep{dekoninck2026matharena}, HumanEval~\citep{chen2021evaluating},
MBPP~\citep{austin2021programsynthesislargelanguage},
LiveCodeBench~\citep{jain2024livecodebenchholisticcontaminationfree}, MT-Bench~\citep{zheng2023judging}, Alpaca~\citep{alpaca}, and
Arena-Hard-v2~\citep{li2024crowdsourceddatahighqualitybenchmarks}, with at most 500 prompts per dataset drawn
with a fixed seed. We apply the Qwen3 chat template with thinking disabled, allow normal EOS termination, and cap
generation at 2,048 new tokens. At concurrency $C$, the client keeps $C$ requests in flight over one queue of
all prompts. The Domino serving check and acceptance experiments use a fixed 20\% per-dataset subset (605 prompts).

\paragraph{Baselines and measurement.}
Markov serving compares client throughput against default SGLang, the same build with the optimized walk
disabled. Both systems run back to back on the same GPU, and runs affected by one-time kernel compilation are
discarded. Both include an upstream SGLang fix for draft-attention boundaries at batch sizes above one, which
affects the 8B drafter. Markov head measurements use the fastest measured SGLang path per configuration.
Recurrent-head references are SGLang's DSpark RNN module, the released DeLS-Spec graph runner, and SGLang's
full-vocabulary Domino rollout with candidate pruning disabled. The Domino serving check uses this same
full-vocabulary baseline.

Head-level measurements execute captured CUDA graphs and include the inputs and outputs required by the caller.
Before each measurement, we flush L2 and refresh the base logits, excluding this preparation from the reported
latency.

\subsection{Serving Speedup}\label{sec:serving}
\begin{figure}[t]
\centering
\includegraphics[width=\linewidth]{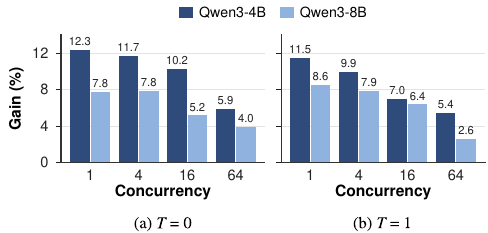}
\caption{End-to-end throughput gains for Qwen3-4B and Qwen3-8B over default SGLang: block-16 Markov drafting
on H100, using all 3,030 prompts.}
\label{fig:serving}
\end{figure}

Figure~\ref{fig:serving} reports throughput with Qwen3-4B-DSpark-b16~\citep{shrey2026dsparkb16}
and Qwen3-8B-Markov-b16~\citep{wang2026xpress}. Gains peak at 12.3\% and 8.6\%, respectively.
The optimized walk improves throughput in
every measured configuration, but the gain decreases at high concurrency. Larger decode batches reduce the
relative head-level advantage, as discussed in
Section~\ref{sec:head-speedup}.

The gain also decreases when the target grows from 4B to 8B. The two configurations use the same vocabulary
size, rank, and walk length, so the optimized head removes a similar amount of work. Target-model computation is
more expensive for 8B, making the saved head time a smaller fraction of end-to-end generation. This single-GPU
trend may reverse under tensor parallelism, where target computation is partitioned but the head remains
replicated on each rank and can occupy a larger share of the critical path.

\subsection{Head-Level Speedup}\label{sec:head-speedup}
\begin{figure*}[t]
\centering
\includegraphics[width=\textwidth]{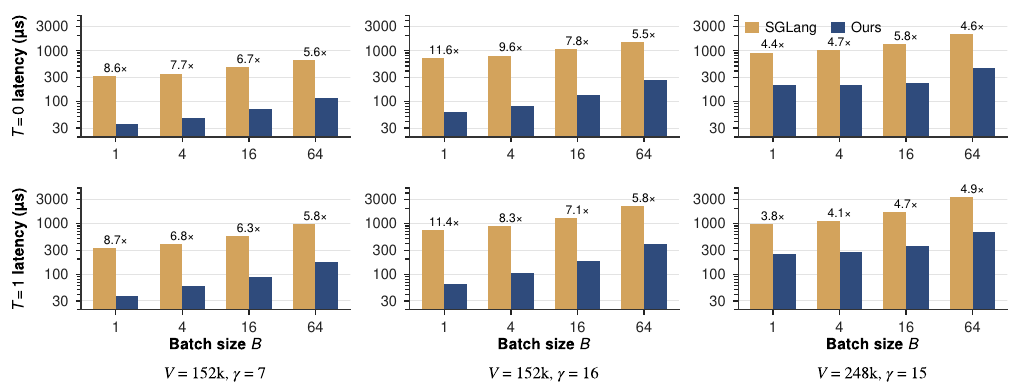}
\caption{Markov-head latency across batch sizes, walk lengths, and vocabularies on H100. Top: $T=0$;
bottom: $T=1$. Latency uses a log scale; labels give speedup over the fastest measured SGLang path.
The 152k path is fully resident at $B\leq4$ and partially resident above; the 248k path is always partially resident.}
\label{fig:walk}
\end{figure*}

Figure~\ref{fig:walk} measures the complete Markov head call while varying batch size, walk length, vocabulary,
and sampling mode. Longer walks reuse the resident projection for more steps and therefore yield larger gains,
reaching \DraftWalkMaxSpeedup$\times$ at $B=1$. Sampling adds epilogue work to both implementations without
changing this trend.

For the standard-vocabulary panels, increasing $B$ has two effects. Accumulators and per-request state leave less
on-chip capacity for resident weights, so more of the projection must be streamed. At the same time, each fetched
weight serves more requests, increasing the arithmetic intensity of the reference projection. Both effects narrow
the relative advantage of residency.

We additionally study the larger-vocabulary head from the Qwen3.6-27B DSpark checkpoint. Its 248,320-row
vocabulary projection does not fit on chip even at $B=1$, so the kernel retains a subset and fetches the rest
at every step. This fixed transfer is about 41\,MB per step, independent of $B$, and limits the $B=1$ speedup to
4.4$\times$ relative to the 127\,MB bf16 reference. As $B$ grows, the transfer is amortized across requests while
the reference's vocabulary-wide per-request work grows, increasing the speedup until computation and selection
costs become significant at the largest batches.

\subsection{Ablations}\label{sec:ablation}
\begin{figure}[t]
\centering
\includegraphics[width=\linewidth]{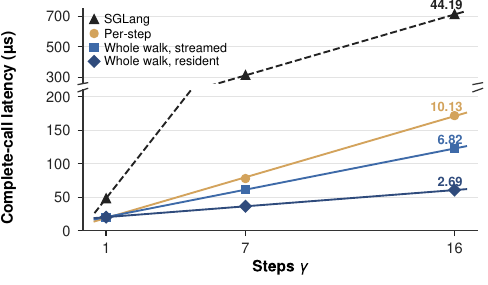}
\caption{Separating fusion and residency gains on H100 ($B=1$, $V=152$k, greedy). Points measure complete
calls; least-squares fits are labeled by their slopes in \textmu s/step. The axis break spans 210--260\us.}
\label{fig:ablation}
\end{figure}

\paragraph{Fusion and residency.}
At $V=151{,}936$ and $B=1$, Figure~\ref{fig:ablation} places three int8 implementations against SGLang: per-step
kernels, a whole-walk kernel that streams the projection each step, and the resident whole-walk kernel. Fusion
reduces the fitted per-step cost by 1.49$\times$; retaining the weights reduces it by another 2.53$\times$. Thus
a single launch alone does not remove the repeated projection traffic.

\begin{figure}[t]
\centering
\includegraphics[width=\linewidth]{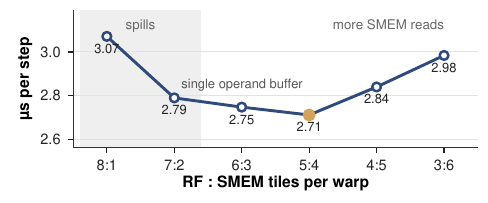}
\caption{RF/SMEM placement tradeoff on H100 ($B=1$, greedy), with 72 weight tiles per CTA. Points are per-step
costs fitted from complete calls at $\gamma=1,7,16$. Shading marks register spills.}
\label{fig:split}
\end{figure}

\paragraph{RF/SMEM placement.}
The fully resident path keeps the same 72 weight tiles per CTA while varying their placement between the
register file and shared memory. More RF tiles eliminate recurring shared-memory operand loads, but leave fewer
registers for execution workspace and eventually cause spills. Figure~\ref{fig:split} shows the resulting
tradeoff. Per-step latency falls as tiles move into RF, reaches a shallow minimum at the 5:4 and 6:3 splits, and
rises at either extreme. At 6:3, the additional RF-resident tiles consume the registers needed to double-buffer
shared-memory operands, so the implementation uses a single buffer.

\subsection{Different Serial Recurrences}\label{sec:head-generality}
\begin{figure}[t]
\centering
\includegraphics[width=\linewidth]{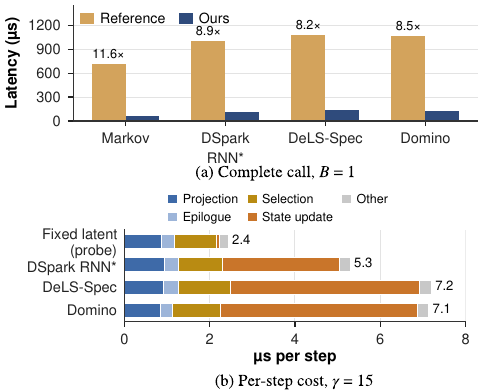}
\caption{Latency and per-step breakdown across serial heads on H100 ($B=1$, greedy).
(a) Complete calls: Markov runs 16 steps, recurrent heads 15; labels give speedup over each head's reference.
*DSpark RNN uses random recurrent weights. (b) Disjoint per-step timings, excluding entry cost.
The fixed-latent probe repeats projection and selection with an unchanged latent.}
\label{fig:heads}
\end{figure}

Figure~\ref{fig:heads} evaluates how the execution structure changes when the selected token advances a recurrent
state. We use released heads for DeLS-Spec and Domino. DSpark defines an RNN recurrence but does not release a
trained checkpoint for it. Our DSpark RNN experiment therefore uses randomly initialized recurrent weights.
The Markov head achieves 11.6$\times$ at $\gamma=16$. At $\gamma=15$, DSpark RNN, DeLS-Spec, and Domino achieve
8.9$\times$, 8.2$\times$, and 8.5$\times$, respectively, against their full-vocabulary references.

Figure~\ref{fig:heads}b shows that projection, epilogue, and token selection remain nearly unchanged across
heads. The added cost follows the number of cross-CTA round trips. The fixed-latent probe communicates across
CTAs only to select a token. Updating the recurrent state adds one gather for DSpark RNN and two for DeLS-Spec
and Domino. Every gather publishes
CTA-local slices, waits for all contributors, and then reads the assembled state through L2, so much of its cost
comes from global synchronization and the L2 round trip rather than recurrent arithmetic. This explains why
DeLS-Spec and Domino have nearly identical per-step latency despite different update equations: both perform
three cross-CTA exchanges per step. Once the vocabulary projection is resident, these exchanges become the main
remaining cost.

As a serving check for a recurrent head, the Qwen3-4B Domino round falls from 6.68 to 5.72\,ms against exact
full-vocabulary SGLang.

\subsection{B200 Capacity Regimes}
\begin{figure}[t]
\centering
\includegraphics[width=\linewidth]{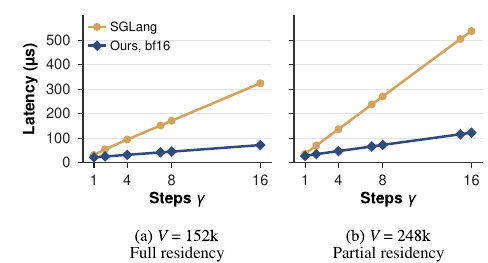}
\caption{B200 bf16 head latency under full (152k) and partial (248k) residency ($B=1$, greedy).}
\label{fig:b200}
\end{figure}

The B200 implementation tests the same organization with bf16 weights and a different set of operand-storage
resources. The 152k projection is fully resident, while the 248k projection keeps about two-thirds of its rows on
chip and streams the remainder. Figure~\ref{fig:b200} separates these capacity regimes.

B200's larger L2 also strengthens the reference path. The 152k projection fits comfortably in L2, allowing
SGLang to reuse it after the first step. At 248k, our kernel streams only the nonresident portion, while SGLang
processes the complete larger projection. Partial residency is therefore slower in absolute time than full
residency, but retains a similar relative gain because the large-vocabulary reference is also slower.

\subsection{Numerical Effects and Acceptance}\label{sec:correctness}
The H100 implementation changes the representation of the vocabulary projection. Consistent with
Figure~\ref{fig:quant}, row-scaled int8 gives 0.58\% Frobenius weight error and 0.26\% projection
error across two released 4B heads~\citep{cheng2026dspark,uiuc2026markov4bb16} and 4,096 sampled latents.
E4M3 with the same per-row scaling gives 2.64\%
and 1.20\%, respectively.

\begin{figure}[t]
\centering
\includegraphics[width=\linewidth]{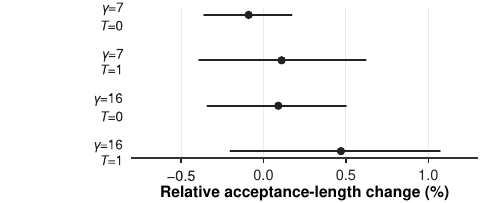}
\caption{Effect of int8 execution on Qwen3-4B draft acceptance length relative to default SGLang. Estimates
pool runs across concurrency settings; lines give 95\% prompt-cluster bootstrap intervals on 605 prompts.}
\label{fig:accept}
\end{figure}

Figure~\ref{fig:accept} shows no systematic acceptance change across the two walk lengths and temperatures. Every
estimate lies near zero, and every confidence interval includes zero. Quantization can therefore change individual
draft proposals without measurably changing acceptance on this workload.

\section{Related Work}\label{sec:related}
\paragraph{Speculative drafting and causal refinement.}
Blockwise decoding~\citep{stern2018blockwise} and speculative sampling~\citep{leviathan2023fast,chen2023accelerating}
reduce the cost of sequential target generation.
Parallel drafting, including DFlash~\citep{chen2026dflash}, shifts computation across proposed positions into a
parallel backbone.
DSpark~\citep{cheng2026dspark}, Domino~\citep{huang2026domino}, and DeLS-Spec~\citep{zheng2026dels} retain
lightweight prefix-dependent refinement.
Earlier dependent-head designs such as Hydra~\citep{ankner2024hydra} and ReDrafter~\citep{cheng2024redrafter}
also expose serial structure, but their weight matrices can exceed the on-chip capacity
targeted by our design.

\paragraph{Persistent execution and weight reuse.}
Persistent RNNs~\citep{diamos2016persistent} established the use of persistent GPU kernels to retain recurrent
weights across time steps. Sparse Persistent RNNs~\citep{zhu2018sparse} and GRNN~\citep{holmes2019grnn} further
explored on-chip recurrent execution.
Our contribution builds on this established idea.
The draft-head setting combines low-rank vocabulary projection with per-step discrete selection and optional
sampling. Its implementation must jointly account for
batch-dependent capacity and distributed state updates.
We evaluate these requirements across multiple head structures and two GPU memory hierarchies.

\paragraph{Megakernels and execution abstractions.}
MPK~\citep{cheng2025mpk} represents tensor programs as SM-level tasks and schedules them within persistent
megakernels; Event Tensor~\citep{jin2026eventtensor} expresses dependencies for dynamic megakernel compilation.
We focus on retaining a weight table over a specific data-dependent recurrence and measuring residency
beyond fusion.

\paragraph{Serving and selection primitives.}
SGLang~\citep{zheng2024sglang} and vLLM~\citep{kwon2023vllm} provide the scheduling and memory-management context
for efficient language-model serving.
Our integration replaces a local draft-head execution path within that context.
GPU sampling kernels in FlashInfer~\citep{ye2025flashinfer} and FlashSampling~\citep{ruiz2026flashsampling}
optimize categorical selection.
Here selection is also the synchronization boundary between dependent steps, so its placement inside the
resident recurrence is central. The optimized head must expose the proposal distribution actually used by
selection to preserve exact speculative sampling.

\section{Conclusion}\label{sec:conclusion}
Serial draft heads combine a small evolving state with a large projection reused across dependent steps.
\sys retains weight shards throughout the walk and advances the recurrence through GPU-side collectives,
supporting Markov and recurrent heads with full or partial residency. Experiments show gains beyond fusion
and higher serving throughput, demonstrating the value of aligning GPU execution with the lifetime of
reusable weights.

\bibliography{refs}
\bibliographystyle{mlsys2025}
\section*{AI Usage Disclosure}

We used Anthropic Claude Code and OpenAI Codex iteratively to assist with literature search and reference
checking, experiment and ablation planning, code implementation and debugging, analysis of measured results,
and manuscript drafting, restructuring, and revision. Code assistance covered GPU kernels, benchmarking
and profiling scripts, and data-analysis and plotting scripts. The tools also assisted with figure preparation
and consistency checks among notation, claims, and reported values. The reported performance results come
from executed GPU experiments. The authors directed the work, retained final control over research decisions,
and take responsibility for the experimental methodology, validation of implementations and measurements,
interpretation of results, references, and final manuscript.

\end{document}